# OPTIMIZED MULTI-LEVEL PRIME ARRAY CONFIGURATIONS

*Saleh A. Alawsh and Ali H. Muqaibel*

Electrical Engineering Department, King Fahd University for Petroleum and Minerals (KFUPM),
Dhahran, 31261 Saudi Arabia,
Email: {salawsh, muqaibel}@kfupm.edu.sa

*ABSTRACT*
Antenna arrays have many applications in direction-of-arrival (DOA) estimation. Sparse arrays such as nested arrays, super nested arrays, and coprime arrays have large degrees of freedom (DOFs). They can estimate large number of sources greater than the number of elements. They also have closed form expressions for antenna locations and the achievable DOFs. The multi-level prime array (MLPA) uses multiple uniform linear subarrays where the number of elements in the subarrays are pairwise coprime integers. The array achieves large DOFs and it has closed form expressions for the antenna locations and the required aperture size. For a given number of subarrays and total number of elements, there are different design alternatives. This paper finds the optimum number of elements within each subarray and the optimized ordered inter-element spacing. In almost all cases, we have found that a unique configuration jointly realizes the maximum number of unique lags and the maximum number of consecutive lags.

## 1. INTRODUCTION

Antenna arrays such as coprime arrays [1] and nested arrays [2] have attracted many researchers because they realize large degrees of freedom (DOFs). Super nested arrays combine the advantages of coprime arrays and nested arrays [3], [4]. Having DOFs greater than the number of antennas of the array is one of the main advantages. These arrays have closed form expressions for antenna locations and the achievable DOFs. Conventional coprime arrays use two uniform linear subarrays (two levels).

In the context of direction of arrival (DOA) estimation, large DOFs using MUSIC algorithm can be realized by maximizing the number of consecutive lags in the difference coarray. While, larger DOFs using compressive sensing (CS) techniques requires exploiting all unique lags. A generalized multi-level prime array (MLPA) was proposed in [5]. The array uses two or more uniform linear subarrays where the number of elements in the subarrays are pairwise coprime integers. There has been some work to find the optimal coprime array configuration [6], [7] which is a special case of the MLPA with just two subarrays. Our aim is to find the optimal configuration for the generalized MLPA.

The rest of the paper is organized as follows. Section 2 presents the signal and system model. Section 3 is dedicated to find the optimum array configurations. Results and discussion are presented in Section 4 and finally Section 5 concludes the paper.

## 2. SIGNAL AND SYSTEM MODEL

A multi-level prime array (MLPA) with $N$ elements uses $L \geq 2$ uniform linear subarrays where the $i^{th}$ subarray has $N_i$ elements for $i = 1,2,\ldots,L$. The variable $L$ also refers to the number of levels. The number of elements in the subarrays are pairwise coprime integers. The MLPA has elements located at [5]:

$$\mathbb{P} = \bigcup_{i=1}^{L} \{k_i \mathcal{S}_i d \mid 0 \leq k_i \leq N_i - 1, \mathcal{S}_i \neq N_i\} \tag{1}$$

where $\mathcal{S}_i d \in \boldsymbol{\mathcal{S}}$ represents the inter-element spacing of the $i^{th}$ subarray with $\boldsymbol{\mathcal{S}} = [\mathcal{S}_1, \mathcal{S}_2, \ldots, \mathcal{S}_L]d$ being the ordered inter–element spacing vector of the subarrays and $\mathcal{S}_i \in \boldsymbol{n} = [N_1, N_2, \ldots, N_L]$. The vector $\boldsymbol{n}$ comprises $L$ pairwise coprime integers. The unit inter-element spacing is $d = \lambda/2$ with $\lambda$ being the wavelength. All subarrays share the location of the first element, so a total of $(L-1)$ elements are repeated which makes the total number of elements, $N$, [5]:

$$N = \sum_{i=1}^{L} N_i - (L - 1) \tag{2}$$

The aperture size of the array can be expressed as [5]:

$$D = \max(\mathcal{S}_{L-1}(N_{L-1} - 1)d, \mathcal{S}_L(N_L - 1)d) \tag{3}$$

We assume that $N_i < N_j$, $\forall i < j$. If we divide the ordered inter-element spacing by $d$ and then sort the entries in an ascending order, we get the number of elements in the subarrays.

## 3. FINDING THE OPTIMAL ARRAYS

There are different design alternatives for the vector $\boldsymbol{n}$ and the ordered inter-element spacing, $\boldsymbol{S}$, for a given $N$ and $L$. We perform exhaustive search to find all possible $\boldsymbol{n}$ vectors that fulfils (2), then we find all possible ordered inter-element spacing. Based on the difference coarray, we extract the number of unique lags, $l_{ug}$, and the number of consecutive lags, $l_{cg}$, where both numbers are function of $\boldsymbol{n}$ and $\boldsymbol{S}$. Let $\boldsymbol{p} = [p_1 d, p_2 d, \dots, p_N d]^T$ be the array element locations with $p_i d \in \mathbb{P}$ for $i = 1, 2, \dots, N$. The difference coarray can be expressed as:

$$\mathbb{D} = \{p_i - p_j | i, j = 1, 2, \dots, N\} \tag{4}$$

where we allow for repetition in $\mathbb{D}$. Given an MLPA with $N$ elements and $L$ subarrays the optimum MLPA can be achieved by either maximizing the number of unique lags or maximizing the number of consecutive lags, which can be formulated as:

$$(\boldsymbol{n}, \boldsymbol{S}) \leftarrow \underset{N_i \in \mathbb{N}^+}{\mathrm{argmax}}\{l_{ug}(\boldsymbol{n}, \boldsymbol{S})\} \tag{5}$$

$$(\boldsymbol{n}, \boldsymbol{S}) \leftarrow \underset{N_i \in \mathbb{N}^+}{\mathrm{argmax}}\{l_{cg}(\boldsymbol{n}, \boldsymbol{S})\} \tag{6}$$

subject to: $N = \sum_{i=1}^{L} N_i - (L-1)$

where $\mathbb{N}^+$ is the set of positive integers.

When the solution of (5) and/or (6) is not unique, other factors like aperture size and reduced mutual coupling can be considered [3]. The mutual coupling is related to $v_\Delta$ which is defined as the number of inter-element spacings, that equals to a unit spacing [8]. The array gets sparser as the value of $v_\Delta$ decreases. If multiple solutions result in the same $v_\Delta$, then the configuration that minimizes aperture size, $D$, is recommended, since this is attractive in antenna designs and array implementations where we have constrains in the physical size.

## 4. RESULTS AND DISCUSSION

The optimum MLPA configurations are constructed for $L = 3, 4, 5$ and $6$. The required inter-element spacing of the subarrays is plotted versus the total number of elements, $N$, in Fig. 1, Fig. 2, Fig. 3, and Fig. 4. There are a total of $2L$ traces in each figure. Half of the traces maximizes the number of unique lags and the other half maximizes the number of consecutive lags. There are cases where (2) cannot be satisfied, like the cases of $N = 11$ for 3LPA and $N = 15$ for 4LPA. This explains the missing values in Fig. 1 and Fig. 2. For the same reason, there is a minimum number of elements after which MLPA configuration can be constructed namely $N = 8, 14, 24$ and $36$ elements for 3LPA, 4LPA, 5LPA and 6LPA, respectively.

As demonstrated in Fig. 1, except for the case with $N = 23$ elements, a unique configuration jointly realizes the maximum $l_{ug}$ and $l_{cg}$. This is demonstrated by overlapping markers. The second subarray is always spaced by $2d$ or $3d$. As a result, the first subarray consists of two or three elements ($N_1 = 2$ or $3$). The first subarray always has the maximum inter-element spacing of $N_3 d$. Consequently, the best ordered inter-element spacing is $\boldsymbol{S}_{3\text{LPA}} = [N_3, N_1, N_2]d$. For the special case of $N = 23$, $\boldsymbol{S}_{3\text{LPA}} = [11, 5, 9]d$ maximizes the unique lags, while the number of consecutive lags is maximized when $\boldsymbol{S}_{3\text{LPA}} = [17, 3, 5]d$.

In most of the 4LPA, the joint optimization of unique and consecutive lags results in a unique design as depicted by the overlapping markers in Fig. 2. There are four cases with multiple solutions highlighted in the figure with arrows. TABLE I summarizes the design alternatives for 4LPA. The case of $N = 14$ or $18$ elements has three design alternatives which jointly optimize the number of unique and consecutive lags as TABLE I illustrates. For $N = 23$ elements, there are three design options to maximize the number of consecutive lags. The best option is when $\boldsymbol{S}_{4\text{LPA}} = [5, 3, 11, 7]d$ which also maximizes $l_{ug}$. Joint optimization cannot be achieved for $N = 31$ elements. As the optimal ordered inter-element spacing for $l_{cg}$ is $\boldsymbol{S} = [5, 19, 3, 7]d$ which is not the same as the two design alternatives illustrated in TABLE I that maximizes $l_{ug}$. In general, the second and the third subarrays have the maximum and the minimum inter-element spacing respectively in most of the considered scenarios. Therefore, the best ordered inter-element spacing is $\boldsymbol{S}_{4\text{LPA}} = [N_2, N_4, N_1, N_3]d$.

In 5LPA, the number of design alternatives increases compared with 4LPA. Fig. 3 specifies that $\boldsymbol{S}_{5\text{LPA}} = [5, 2, 17, 3, 13]d$ maximizes $l_{ug}$, while the number of consecutive lags is maximized when $\boldsymbol{S}_{5\text{LPA}} = [5, 2, 19, 3, 11]d$ when $N = 36$ elements. In Table II, we present only the design alternatives that maximize either $l_{ug}$ or $l_{cg}$. For $N = 30$ and $44$ elements, one of those alternatives also jointly maximizes the number of consecutive lags as Fig. 3 and Table II confirm. Joint optimization cannot be achieved for $N = 32$ and $41$ elements.

Others jointly maximize both lags are not included in the table because of the space. For example, the case of $N = 28$ elements has six design alternatives which jointly maximize both lags. In addition, two design alternatives at $N = 26,35$ elements and three design alternatives when $N = 24$ elements are jointly improve both lags. In Fig. 3 and Table II, most of the investigated scenarios have ordered inter-element spacing of $\boldsymbol{S}_{5LPA} = [N_3, N_1, N_5, N_2, N_4]d$ or $\boldsymbol{S}_{5LPA} = [N_3, N_1, N_2, N_5, N_4]d$.

Larger MLPA levels require large number of elements, 36 in case of 6LPA. The 6LPA with $N = 42$ elements attains its maximum $l_{ug}$ with $\boldsymbol{S} = [7,2,3,5,17,13]d$, see Fig. 4. On the other hand, it has got two ordered inter-element spacings that maximize $l_{cg}$ as Table III depicts namely: $\boldsymbol{S}_1 = [7,2,3,5,19,11]d$ and $\boldsymbol{S}_2 = [5,2,7,3,19,11]d$ based on different $\boldsymbol{n}$.

In Fig. 5, the number of unit spacings, $v_\Delta$, for the 3LPA and 4LPA configurations is plotted as a function of the total number of elements. Nested arrays, super nested arrays, and coprime arrays are included for comparison. The value of $v_\Delta$ is either 3 or 4 for the 3LPA regardless of the total number of elements. For the special case of $N = 23$ elements, the configurations that maximizes the number of unique lags has $v_\Delta = 4$, while the other that maximizes the number of consecutive lags has $v_\Delta = 3$. In 4LPA, $v_\Delta$ is oscillating between 5 and 9. There are some design alternatives that result in equal $v_\Delta$ when $N = 18$ and 23 elements. While, others generate different $v_\Delta$ values as in $N = 14$ and 31 elements.

In comparison, coprime arrays have two pairs of elements spaced by half wavelength, so $v_\Delta = 2$. The value of $v_\Delta$ increases with the number of elements $N$ in nested arrays [3], [4]. On the other hand, there is a maximum of two unit spacings for super nested arrays. The later was mainly proposed to mitigate the mutual coupling effect in nested arrays. The value of $v_\Delta$ in 3LPA is not very far from that of coprime array. In addition, the 3LPA has fewer holes and it requires smaller aperture size [9]. This is attractive in antenna designs and array implementations where we have constrains in the physical size [5].

## 5. CONCLUSION

Two different optimization criteria were used to design MLPA configurations. In most of the cases, there is a unique solution which jointly maximizes the number of unique lags and the number of consecutive lags which makes the optimum MLPA configuration unique. The number of design alternatives increases as the level of the array increases. In this case, multiple solutions realize the maximum number of lags and other factors like reduced mutual coupling can be used to make the final selection.

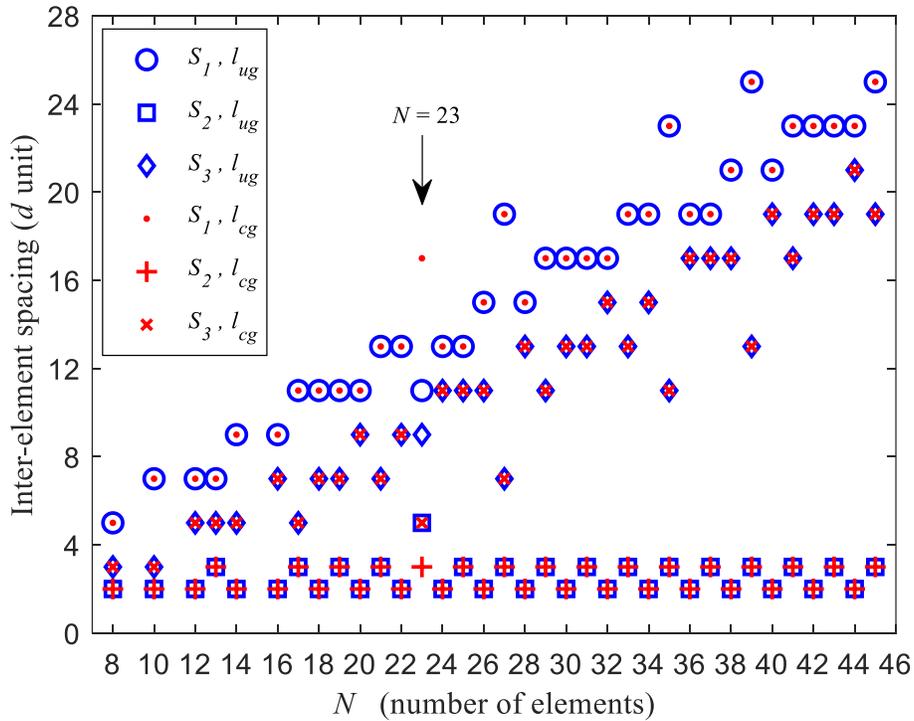

**Fig. 1.** The optimal inter-element spacing versus the number of elements for 3LPA ( $L = 3$ )

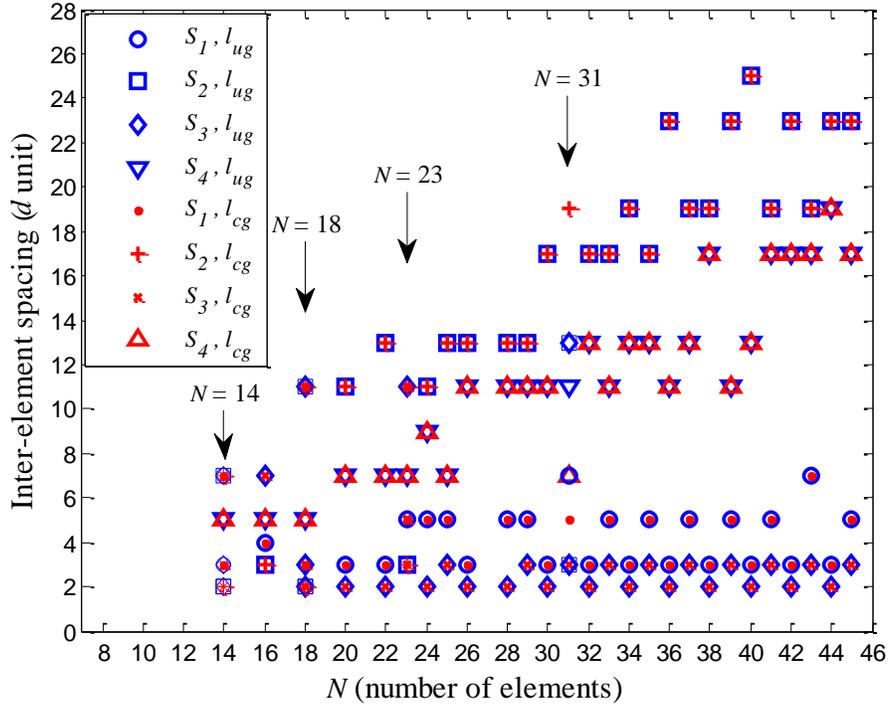

**Fig. 2.** The optimal inter-element spacing versus the number of elements for 4LPA ($L = 4$)

**TABLE I.** Design alternatives for 4LPA

| $S \over N$ | $S_1$ | $S_2$ | $S_3$ | $S_4$ | Lags |
|---|---|---|---|---|---|
| 14 | 7 | 2 | 3 | 5 | |
|    | 3 | 7 | 2 | 5 | |
|    | 3 | 2 | 7 | 5 | $l_{ug}, l_{cg}$ |
| 18 | 11 | 2 | 3 | 5 | |
|    | 3 | 11 | 2 | 5 | |
|    | 3 | 2 | 11 | 5 | |
| 23 | 11 | 3 | 5 | 7 | |
|    | 5 | 11 | 3 | 7 | $l_{cg}$ |
|    | **5** | **3** | **11** | **7** | |
| 31 | 7 | 13 | 3 | 11 | $l_{ug}$ |
|    | 7 | 3 | 13 | 11 | |

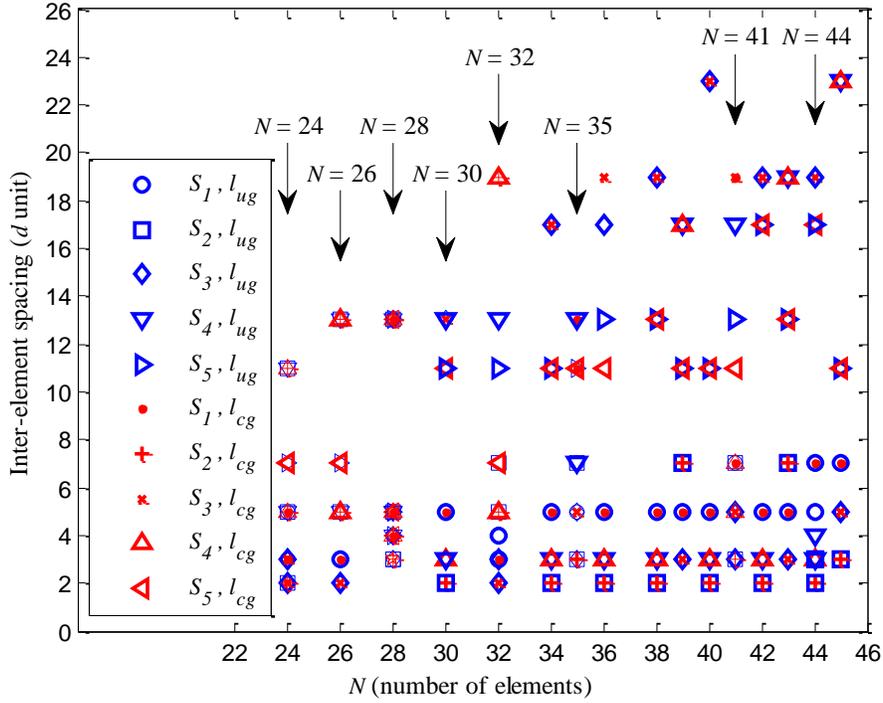

**Fig. 3.** The optimal inter-element spacing versus the number of elements for 5LPA ( $L = 5$ )

**Table II.** Design alternatives for 5LPA

| $\dfrac{S}{N}$ | $S_1$ | $S_2$ | $S_3$ | $S_4$ | $S_5$ | Lags |
|---|---|---|---|---|---|---|
| 24 | 5 | 2 | 3 | 11 | 7 | |
|    | 3 | 5 | 2 | 11 | 7 | |
|    | 3 | 11 | 2 | 5 | 7 | |
| 26 | 3 | 5 | 2 | 13 | 7 | |
|    | 3 | 13 | 2 | 5 | 7 | |
|    | 13 | 3 | 4 | 5 | 7 | $l_{ug}, l_{cg}$ |
|    | 5 | 3 | 13 | 4 | 7 | |
| 28 | 5 | 3 | 4 | 13 | 7 | |
|    | 4 | 5 | 3 | 13 | 7 | |
|    | 4 | 13 | 3 | 5 | 7 | |
|    | 4 | 3 | 13 | 5 | 7 | |
| 30 | 5 | 2 | 13 | 3 | 11 | $l_{ug}$ |
|    | 5 | 2 | 3 | 13 | 11 | |
| 32 | 3 | 7 | 2 | 13 | 11 | $l_{ug}$ |
|    | 4 | 5 | 3 | 13 | 11 | |
|    | 3 | 5 | 2 | 19 | 7 | $l_{cg}$ |
|    | 3 | 19 | 2 | 5 | 7 | |
| 35 | 7 | 3 | 5 | 13 | 11 | $l_{ug}, l_{cg}$ |
|    | 5 | 7 | 3 | 13 | 11 | |
| 41 | 7 | 3 | 5 | 17 | 13 | $l_{ug}$ |
|    | 5 | 7 | 3 | 17 | 13 | |
|    | 19 | 3 | 5 | 7 | 11 | |
|    | 7 | 3 | 19 | 5 | 11 | $l_{cg}$ |
|    | 5 | 19 | 3 | 7 | 11 | |
| 44 | 7 | 2 | 19 | 3 | 17 | $l_{ug}$ |
|    | 5 | 3 | 19 | 4 | 17 | |

**Table III.** Design alternatives for 6LPA

| $S$ \ $N$ | $S_1$ | $S_2$ | $S_3$ | $S_4$ | $S_5$ | $S_6$ | Lags |
|---|---|---|---|---|---|---|---|
| 36 | 7 | 2 | 3 | 5 | 13 | 11 | $l_{ug}, l_{cg}$ |
|    | 3 | 7 | 2 | 5 | 13 | 11 |  |
| 38 | 7 | 3 | 4 | 5 | 13 | 11 | $l_{ug}, l_{cg}$ |
|    | 5 | 3 | 7 | 4 | 13 | 11 |  |
|    | 4 | 7 | 3 | 5 | 13 | 11 |  |
|    | 4 | 3 | 7 | 5 | 13 | 11 |  |
| 42 | 7 | 2 | 3 | 5 | 19 | 11 | $l_{cg}$ |
|    | 5 | 2 | 7 | 3 | 19 | 11 |  |

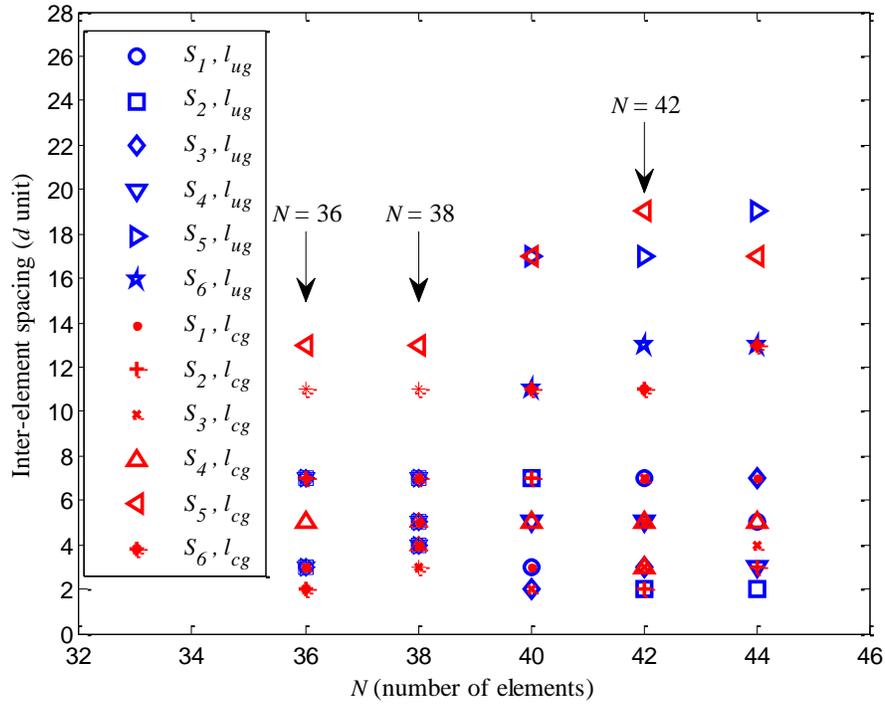

**Fig. 4.** The optimal inter-element spacing versus the number of elements for 6LPA ($L = 6$)

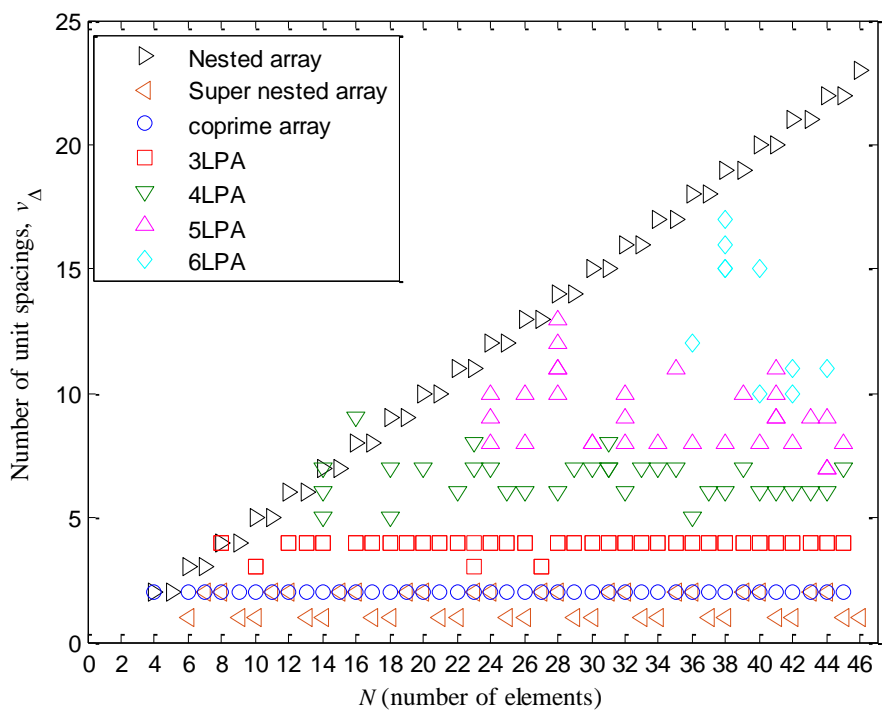

**Fig. 5.** Number of unit spacings versus the total number of elements

## 6. ACKNOWLEDGMENT

This work was supported by the Deanship of Scientific Research (DSR) at King Fahd University of Petroleum & Minerals (KFUPM), Dhahran, Saudi Arabia, under Project No. IN161015.

## 7. DATA AVAILABILITY STATEMENTS

The datasets generated and/or analysed during the current study are not publicly available due to so many Matlab files (.mat) that are scattered here and there and due to the different options available to build a certain array configuration but are available from the corresponding author on reasonable request.

## 8. REFERENCES

[1] P. P. Vaidyanathan and P. Pal, "Sparse Sensing with Co-prime Samplers and Arrays," *IEEE Trans. Signal Process.*, vol. 59, no. 2, pp. 573–586, 2011.
[2] P. Pal and P. P. Vaidyanathan, "Nested Arrays: A novel Approach to Array Processing with Enhanced Degrees of Freedom," *IEEE Trans. Signal Process.*, vol. 58, no. 8, pp. 4167–4181, 2010.
[3] C. L. Liu and P. P. Vaidyanathan, "Super Nested Arrays: Linear Sparse Arrays With Reduced Mutual Coupling-Part I: Fundamentals," *IEEE Trans. Signal Process.*, vol. 64, no. 15, pp. 3997–4012, 2016.
[4] C. L. Liu and P. P. Vaidyanathan, "Super Nested Arrays: Linear Sparse Arrays With Reduced Mutual Coupling—Part II: High-Order Extensions," *IEEE Trans. Signal Process.*, vol. 64, no. 16, pp. 4203–4217, 2016.
[5] Saleh A. Alawsh and Ali H. Muqaibel, "Multi-Level Prime Array for Sparse Sampling," *IET Signal Process.*, vol. 12, no. 6, pp. 688 – 699, Aug. 2018.
[6] K. Adhikari, J. R. Buck, and K. E. Wage, "Beamforming with Extended Co-prime Sensor Arrays," *IEEE Int. Conf. Acoust. Speech Signal Process. (ICASSP), Vancouver, BC, Canada*, pp. 4183–4186, May 2013.
[7] K. Adhikari, J. R. Buck, and K. E. Wage, "Extending Coprime Sensor Arrays to Achieve the Peak Side Lobe Height of a Full Uniform Linear Array," *EURASIP J. Adv. Signal Process.*, vol. 2014, no. 148, pp. 1–17, 2014.
[8] R. Rajamaki and V. Koivunen, "Comparison of Sparse Sensor Array Configurations with Constrained Aperture for Passive Sensing,"

*IEEE Radar Conf. (RadarConf), Seattle, WA, USA*, no. 1, pp. 0797–0802, May 2017.

[9] S. A. Alawsh and A. H. Muqaibel, "Three-Level Prime Arrays for Sparse Sampling in Direction of Arrival Estimation," *IEEE Asia-Pacific Conf. Appl. Electromag (APACE), Langkawi, Malaysia*, pp. 277–281, Dec. 2016.